\documentclass[aps,prx,twocolumn,showpacs,citeautoscript,superscriptaddress,longbibliography,notitlepage]{revtex4-1}

\usepackage{graphicx}  % needed for figures
\usepackage{epstopdf}
\usepackage{dcolumn}   % needed for some tables
\usepackage{bm}        % for math
\usepackage{amsmath}
\usepackage{amssymb}   % for math
\usepackage{wrapfig}
\usepackage{array}
\usepackage[caption=false]{subfig}
\usepackage[bookmarks=false,linkcolor=blue,urlcolor=blue,colorlinks,citecolor=blue]{hyperref}
\usepackage{exscale,relsize}
\usepackage[usenames, dvipsnames]{color}
\usepackage{times, verbatim, color, ulem}
\usepackage{pstricks}
\usepackage{bbold}

\usepackage{pst-all}

\newcommand{\bra}[1]{\langle #1|}
\newcommand{\ket}[1]{|#1\rangle}

\makeatletter

\begin{document}

\title{Fractional Chern insulator edges and layer-resolved lattice contacts}

\author{Christina Knapp}
\affiliation{Department of Physics, University of California, Santa Barbara,
	California 93106 USA}

\author{Eric M. Spanton}
\affiliation{California Nanosystems Institute, University of California, Santa Barbara,
	California 93106 USA}

\author{Andrea F. Young}
\affiliation{Department of Physics, University of California, Santa Barbara,
	California 93106 USA}

\author{Chetan Nayak}
\affiliation{Station Q, Microsoft Research, Santa Barbara, California 93106-6105, USA}
\affiliation{Department of Physics, University of California, Santa Barbara,
	California 93106 USA}

\author{Michael P. Zaletel}
\affiliation{Department of Physics, University of California, Berkeley,
	California 94720 USA}
\affiliation{Department of Physics, Princeton University, Princeton, New Jersey 08540 USA}

\begin{abstract}
Fractional Chern insulators (FCIs) realized in fractional quantum Hall systems subject to a periodic potential are topological phases of matter for which space group symmetries play an important role.  In particular, lattice dislocations in an FCI can host non-Abelian topological defects, known as genons.  Genons can increase the ground state degeneracy of the system and are thus potentially useful for topological quantum computing.   In this work, we study FCI edges and how they can be used to detect genons.  We find that translation symmetry can impose a quantized momentum difference between the edge electrons of a partially-filled Chern band.  We propose {\it layer-resolved lattice contacts}, which utilize this momentum difference to selectively contact a particular FCI edge electron.  The relative current between FCI edge electrons can then be used to detect the presence of genons in the bulk FCI.  Recent experiments have demonstrated graphene is a viable platform to study FCI physics.  We describe how the lattice contacts proposed here could be implemented in graphene subject to an artificial lattice, thereby outlining a path forward for experimental dectection of non-Abelian topological defects.
\end{abstract}
\date{\today}
\maketitle

{\it Introduction}. Non-Abelian topological physics has excited intense interest in the condensed matter community, in part for its potential application to quantum computing~\cite{Kitaev03,Nayak08}.  Traditionally, the emphasis has been to discover non-Abelian topological phases, whose emergent quasiparticles are {\it non-Abelian anyons}.  Non-Abelian anyons have an internal degenerate state space that can encode quantum information, and satisfy exotic braiding statistics such that their adiabatic exchange can result in a unitary rotation within the ground state subspace.  While more than three decades of searching for non-Abelian anyons has resulted in some progress~\cite{Dolev08,Radu08,Willett10,Banerjee17}, it has also emphasized the difficulty of conducting such experiments.  An attractive alternative is to engineer extrinsic defects with non-Abelian braiding statistics and ground state degeneracy.  Such topological defects are potentially more experimentally manageable because their location and number can be controlled.
Majorana zero modes (MZMs) in topological superconductors~\cite{Kitaev01}, defects with Ising anyon fusion and braiding statistics, have been the focus of these studies due to their relative experimental accessibility~\cite{Lutchyn10,Oreg10,Alicea12,Mourik12,Rokhinson12,Finck12,Das12,Deng12,Churchill13,NadjPerge14,Albrecht16,Zhang18,Lutchyn18}.  Unfortunately, the braiding statistics of MZMs do not support universal quantum computation~\cite{Gottesman98a}, thus most Majorana-based quantum computing proposals rely on resource-expensive distillation protocols~\cite{Bravyi05,Bravyi06,Bravyi12,Duclos12,Haah17a,Haah17b}. 
It therefore remains desirable to engineer alternative, more computationally powerful topological defects.

One potential alternative are genons- topological defects whose presence effectively changes the genus of the sytem~\cite{Barkeshli12,Barkeshli15,Barkeshli16,Cong17}.  Genons can increase the ground state degeneracy of an otherwise Abelian topological phase, enhancing the computational power of the system.  One system in which genons are predicted to appear are fractional Chern insulators (FCIs)~\cite{Haldane88,Regnault11,Neupert11,Parameswaran13,Bergholtz13,Moller15}.  An FCI is a topological phase occuring at partial filling of a band with non-trivial Chern number $C\in \mathbb{Z}/\{0\}$.  The fractional quantum Hall (FQH) effect is a special case of an FCI, in which all bands (Landau levels) have $C=1$.  Applying a periodic potential ({\it e.g.} a lattice) to a QH system can result in bands with $|C|>1$.  The ground state of a partially filled Chern-$C$ band can be mapped to a $|C|$-layer FQH state in which different lattice sites are analogous to layers~\cite{Qi11,Wu12,Barkeshli13,Harper14,Ippoliti18}.  Lattice symmetries are thus interwoven with internal component labels of the FCI; translations have a non-trivial action on layer index which can result in genons localized at lattice dislocations~\cite{Barkeshli13}.  

Recent experiments have demonstrated that FCIs can be realized in graphene, where the periodic potential arises from a Moir\'e pattern formed by interference between the graphene and dielectric lattices~\cite{Spanton17}.  These experiments indicate that graphene is a viable platform in which to pursue non-Abelian physics, however the Moir\'e potential is not readily applicable to genons as it is difficult to controllably insert lattice dislocations into the Moir\'e superlattice.
 Alternatively, the lattice potential can be engineered, {\it e.g.}, by patterning holes into a neighboring metallic gate or dielectric~\cite{Albrecht01,Hensgens17,Forsythe17}.  An artificial lattice is an appealing route towards realizing FCIs hosting genons because (1) the lattice itself can be used to tune to the desired phase, and (2) there is no additional cost associated with patterning dislocations.

\begin{figure}[t] 
\begin{center}
	\includegraphics[width=\columnwidth]{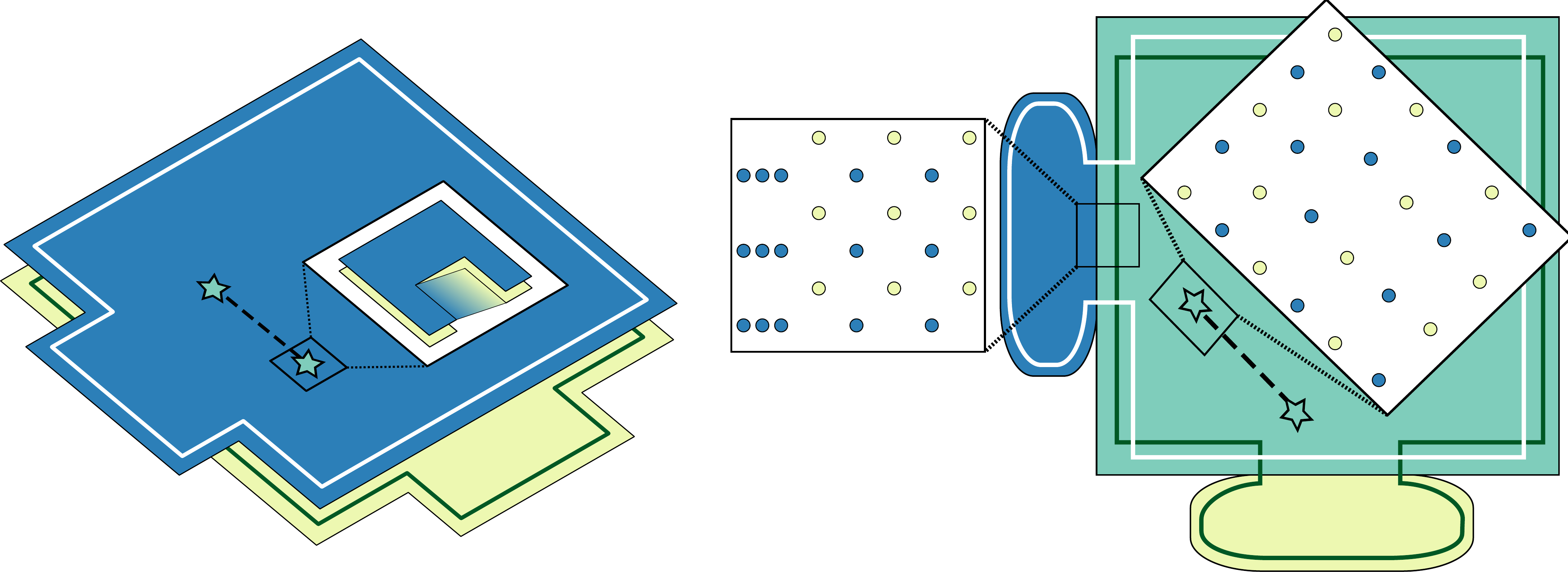}
	\caption{
Analogy between a bilayer QH system ({\it left panel}) and an FCI in a $C=2$ band ({\it right panel}).  Both systems contain a pair of genons (stars) and a blue and yellow region to selectively contact the two edge electrons (white and green lines).  A genon in the bilayer system exchanges the layers.  {\it Right panel.}  An FCI (green) with two layer-resolved lattice contacts (blue and yellow).  Each lattice contact gaps out one of the FCI's edge electrons, rerouting that electron along the exterior of the contact and allowing for selective voltage-bias and current measurement.  The FCI is in a partially-filled $C=2$ band subject to a square lattice potential, such that it realizes the two-component $(mml)$ phase.  The two components, `layers', are localized on the blue and yellow sublattices.  The contacts are in a $C=1$ band.  The unit cell area of the rectangular lattice is half that of the bulk, and is lattice-matched with the bulk along the interface.
}
	\label{fig:contacts}
\end{center}
\end{figure}

Even after genons have been engineered, there remains a final hurdle of how to detect their presence, which is the focus of this work.  To understand why this is challenging, it is helpful to consider the analogy of an FCI in a Chern-2 band to a bilayer QH system, depicted in Fig.~\ref{fig:contacts}.   When the FCI ground state satisfies the microscopic lattice symmetries, sublattices are analogous to layers.  Crucially, under this mapping unit cell translations and plaquette-centered rotations interchange the two sublattices, therefore lattice dislocations play the same role as layer-exchange defects in the bilayer system (left-most inset).  In the bilayer case, layer-exchange defects can be detected using the difference in the edge current of the two layers~\cite{Barkeshli14}, which in turn can be measured by separately contacting each layer's edge.
In the FCI case, the difference in the current associated with the edge electrons again carries a signature of the genon; however, we must devise a way to selectively contact edge electrons residing in the same physical sample.

In this work, we study FCI edges in a partially filled $C>1$ band and propose {\it layer-resolved lattice contacts} that can be used to detect genons.  The main idea is depicted in the left panel of Fig.~\ref{fig:contacts}.  Essentially, a local translation symmetry along the edge constrains the allowed perturbations from electon tunneling between the FCI (green) and lattice contacts (blue and yellow).  By appropriately designing the lattices in the three regions, the two contact interfaces can gap out different edge electrons of the FCI, thereby spatially separating them and allowing independent measurement of their electrical properties.  The relative current can then be used to detect genons in the bulk~\cite{Barkeshli14}, providing a path forward for experimental detection of non-Abelian topological defects in graphene.  

The remainder of this paper is organized as follows.  We briefly review the mapping of an FCI ground state to a $|C|$-layer QH state.  We next study the FCI edge physics, elucidating the additional constraints translation symmetry imposes on electron tunneling across the interface.  We then discuss how the lattice itself can be used as a tuning parameter to simultaneously realize different phases in the same sample.  Finally, we synthesize the above discussion to propose layer-resolved lattice contacts and illustrate how these contacts provide the missing link in experimental dectection of genons.  

%%%  
{\it Preliminaries.} Consider a square lattice with unit cell area $a^2$ and perpendicular magnetic field such that the flux density is $\phi=p/q$ with $p$ and $q$ coprime integers.  
Chern bands are characterized by topological invariants $C$ and $S$ given by theTKNN Diophantine equation~\cite{Thouless82}
\begin{equation}\label{eq:TKNN}
n_e=C \phi +S,
\end{equation}
where $n_e$ is the electron density per unit cell. 
In Appendix~\ref{app:mapping}, we review how the single particle orbitals of the band can be mapped to a $|C|$-layer QH system at flux density $\bar{\phi}=\phi+S/C$ with effective magnetic length ${\bar{\ell}_B=a/\sqrt{2\pi \bar{\phi}}}$.  
Recall that in the Landau gauge ${\mathbf{A} = B (-y, 0)}$ of a continuum Landau level, single-particle states are uniquely labeled by their momentum $k_x$.
The key point is that in an appropriate basis, single particle orbitals $\ket{\tilde{k}_x, \beta}$ of a Chern-$C$ band have a continuum index $\tilde{k}_x\in \mathbb{R}$ analogous to this momentum, and an internal index $\beta\in \mathbb{Z}_C$ analogous to ``layer.''  Translations and $C_4$ rotations factor into continuum and internal parts $T_j=\tilde{T}_j\otimes \tau_j$, $j=x,y$; $C_{4,l}=\tilde{C}_{4,l}\otimes \gamma_{4,l}$, $l=p,s$ denoting plaquette-centered and site-centered rotations, respectively ($C_{4,s}=T_x C_{4,p})$.  The continuum parts, denoted with a tilde, transform $\tilde{k}_x$ just as in a continuum Landau level at flux density $\bar{\phi}$.  When $S$ and $C$ are coprime, the internal parts, denoted with a greek letter, act non-trivially on the layer index: $\tau_x \tau_y = e^{2 \pi i S / C} \tau_y\tau_x   $. 
 In the limit that $\bar{\phi}\to 0,$ the system has a continuum limit and admits a field theoretic description. 
This is the precise sense in which a Chern-$C$ band is like a $|C|$-layer QH system, with lattice symmetries acting as internal symmetries on the layer index~\cite{Qi11,Wu12,Barkeshli13,Harper14,Ippoliti18}.

For concreteness, we consider a partially filled $C=2$, $S$ odd band whose ground state realizes an Abelian, $C_4$-symmetric $(mml)$ state.  
At the topological level the system is described by the Lagrangian
\begin{align}\label{eq:L-b}
\mathcal{L} = \frac{1}{4\pi} \int dx \left\{ K_{IJ} a_{I,\mu}\partial_\nu a_{J,\nu}\varepsilon^{\mu \nu \lambda} + 2 t_I A_\mu \partial_\nu a_{I,\lambda} \varepsilon^{\mu \nu \lambda}\right\},
\end{align}
where $K_{IJ}$ is a $2\times 2$ universal matrix describing the phase, $\mathbf{t}=(1,1)$ is the charge vector, $\mathbf{a}_J$ are the Chern-Simons gauge fields, and $\mathbf{A}$ is the external electromagnetic vector potential.  
The topological field theory must then be supplemented with the symmetry action. 
The electron current in layer $I$ is ${j^{\mu}_{e, I} = \frac{1}{2\pi}\partial_\nu a_{I,\lambda} \varepsilon^{\mu \nu \lambda}}$, while the electron operator $\psi_{e, I}$ generates a corresponding flux in $\mathbf{a}_I$. 
We demand that the $\psi_{e, I}$ transform under the lattice symmetries just like the single-particle orbitals of a $C=2$ band; specifically they transform under translations as $\tau_j=\sigma_j$
where $\sigma_{x/y}$ are Pauli matrices, and under rotations as $\gamma_{4,p}=(\tau_x+\tau_y)/\sqrt{2}$. This implicitly defines the action of the symmetry on the Chern-Simons fields, as detailed below.
Note that by a change of basis in the layer space $\beta$, we could have instead chosen (say) $\tau_x = \sigma_z$; this corresponds to a {\it distinct} implementation of the symmetry (in fact, it corresponds to a ``topological nematic'' state~\cite{Barkeshli12}). Our choice is $C_4$ symmetric.
When $|m-l|\geq 2$, interchanging the layers permutes the anyons, and consequently~\cite{Cheng16} such twists defects are genons with quantum dimension $d=\sqrt{|m-l|}$~\cite{Barkeshli12,Barkeshli13}.
For our choice $\tau_{j} = \sigma_j$, a lattice dislocation with a Burger's vector along {\it either} $x$ or $y$ permutes the layers, so will carry this degeneracy.

%%%
{\it FCI edge states}. 
The interplay of translation symmetry and the component labels of the many-body state has interesting implications for FCI edge states.  The Lagrangian associated with the edge of the system is~\cite{Wen92} 
\begin{align}\label{eq:L-edge}
\mathcal{L}_\text{edge}= \frac{1}{4\pi} \int dx \left\{ K_{IJ} \partial_t \phi_I \partial_x \phi_J - V_{IJ}\partial_x \phi_I \partial_x \phi_J\right\},
\end{align}
where the matrix $K_{IJ}$ is that of the bulk theory, while the edge potential $V_{IJ}$ is non-universal.
We continue to work in the basis with $\mathbf{t}=(1,1)$, so that $I$ and $J$ are layer indices.
For concreteness, consider an edge along the $(u,v)$ direction with translation symmetry ${T_{(u,v)}=\tilde{T}_{(u,v)}\otimes \tau_{(u,v)}}$, {\it e.g.}, $(u,v)=(1,0)$ corresponds to an $x$-edge.  The ``internal'' part of the translation acts on the electron operators $\psi_{e,I}\sim \text{exp}\left\{ i K_{IJ} \phi_J \right\}$ as $\tau_{(u,v)} \left(\psi_{e,1}, \psi_{e,2}\right)^T$.  
This implies, for example, that $x$ translation interchanges the bosonic edge modes for an $(mml)$ phase: $\tau_x\phi_{1/2}=\phi_{2/1}$.

Translation symmetry imposes additional constraints on the allowed perturbations to the FCI edge theory.  Consider a translationally invariant interface between two phases described by $K^L$ and $K^R$.  When these phases are not related by anyon condensation, only perturbations arising from electron tunneling across the interface are allowed.  These perturbations take the form
\begin{align}\label{eq:cosine}
t_{gh} \cos\left( g_I K_{IJ}^L \phi_J^L - h_I K_{IJ}^R \phi_J^R\right),
\end{align}
where $g_I$ and $h_I$ are integer vectors satisfying ${\sum_I g_I=\sum_I h_I}$ from charge conservation.  In the continuum QH effect, Eq.~\eqref{eq:cosine} can gap out the edge modes $e^{i g_I K_{IJ}^L \phi_J^L}, e^{i h_I K_{IJ}^R \phi_J^R}$ when the left and right scaling dimensions are equal ($g_I K_{IJ}^L g_J^T= h_I K_{IJ}^R h_J^T$) and the total scaling dimension is less than two.  For the FCI interface considered here, Eq.~\eqref{eq:cosine} must additionally be invariant under the component translation symmetries of the left and right phases, $\tau_{(u,v)}^{L/R}$. 

\begin{figure}[t]	
	\includegraphics[width=\columnwidth]{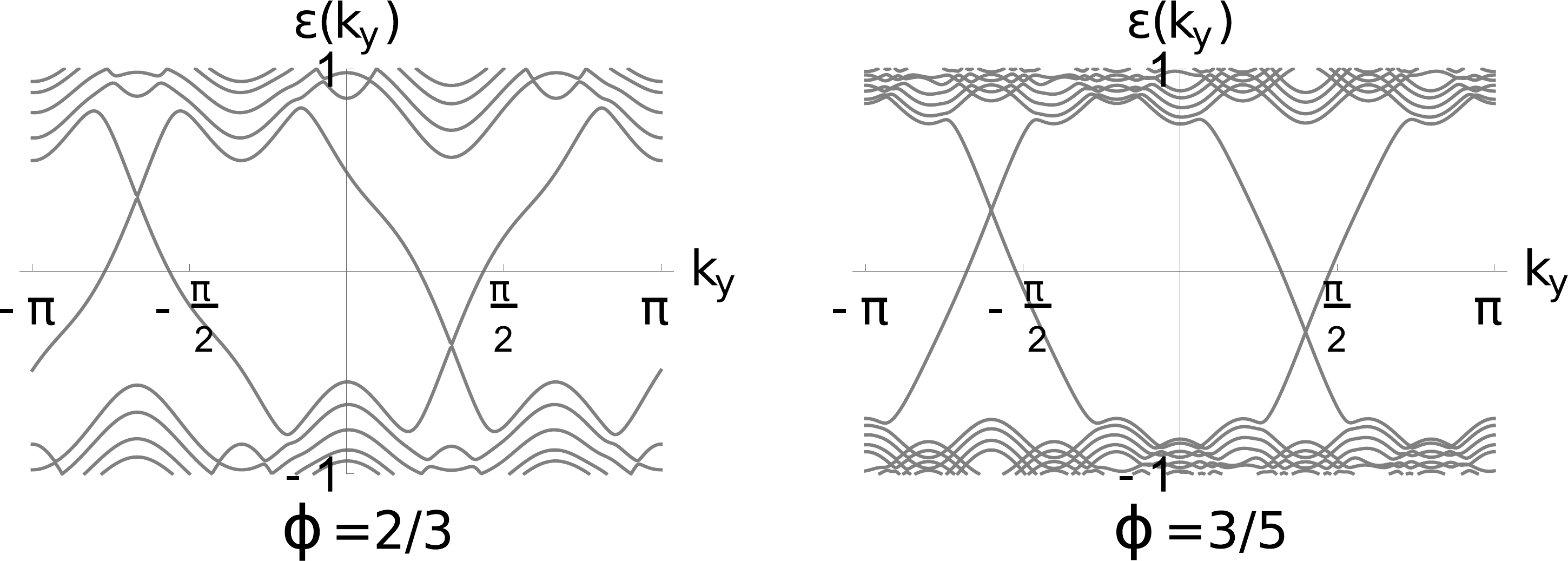}
	\caption{Edge states for the Hofstadter model near $\phi=1/2$ on the infinite cylinder.  The two halves of the cylinder differ by filling a band with $C=2$, $S=-1$, with left and right movers corresponding to opposite edges.  
The edge state momentum difference at $\varepsilon=0$ is $\frac{\pi}{a}(1+1/q)$ for flux density $p/q$.  As $p/q\to 1/2$, the edge state momentum difference approaches quantization, corresponding to the limit that the system admits a field theory description (see Appendix~\ref{app:edge-details}).
 }\label{fig:Hof1}
\end{figure}

An alternative way of understanding this additional constraint is that the interplay of translation symmetry and layer index introduces a difference in the edge momenta that is not present for the analogous FQH state.
Consider the $(mml)$ state with $\tau_{x/y}=\sigma_{x/y}$.  An edge along the $(1,1)$ direction has $\tau_{(1,1)}= \sigma_z$.  When $V_{IJ}$ is a symmetric matrix, the layer-exchange symmetry implies that for a bilayer FQH state we would expect the edge electrons associated to the two layers to have the same momenta; this implies that for an FCI, both edge electrons $\psi_{e,1/2}$ have the same $\tilde{k}_{(1,1)}$.  However, the {\it internal} part of the translation introduces a quantized momentum difference of $\pi/\left(\sqrt{2}a\right)$:
\begin{align}\label{eq:edge-momenta}
\tilde{T}_{(1,1)}\otimes \tau_{(1,1)} \left( \begin{array}{c} \psi_{e,1} \\ \psi_{e,2} \end{array}\right)& = e^{i \tilde{k}_{(1,1)} \sqrt{2}a }  \left( \begin{array}{c} \psi_{e,1} \\ - \psi_{e,2} \end{array}\right).
\end{align}
This momentum difference will no longer be quantized in the presence of disorder or non-symmetric perturbations to $V_{IJ}$, however, provided these are small effects, the edge momenta will be roughly separated by $\pi$ over the lattice spacing along the edge.  In Fig.~\ref{fig:Hof1}, we plot this momentum difference for the Hofstadter model and show that it approaches quantization in the limit $\phi\to -S/C$~\cite{Hofstadter76,Harper55}.  

When the system satisfies plaquette-centered $C_4$ symmetry, the $(1,-1)$ edge has component translation ${\tau_{(1,-1)}={\gamma_{4,p}^{-1} \tau_{(1,1)} \gamma_{4,p}=-\sigma_z}}$,
and the momenta of the edge electrons are swapped compared to Eq.~\eqref{eq:edge-momenta} (assuming $\tilde{k}_{(1,1)}=\tilde{k}_{(1,\pm 1)}$).  For the $(331)$ state, there is an MZM at the corner, which interchanges the two layers of the FCI (see Appendix~\ref{app:MZMs}).  Due to the presence of gapless edge modes, these corner MZMs are not exponentially localized the way that topological defects in the gapped bulk are.  

\begin{figure}[t]
\begin{center}
	\includegraphics[width=.3\columnwidth]{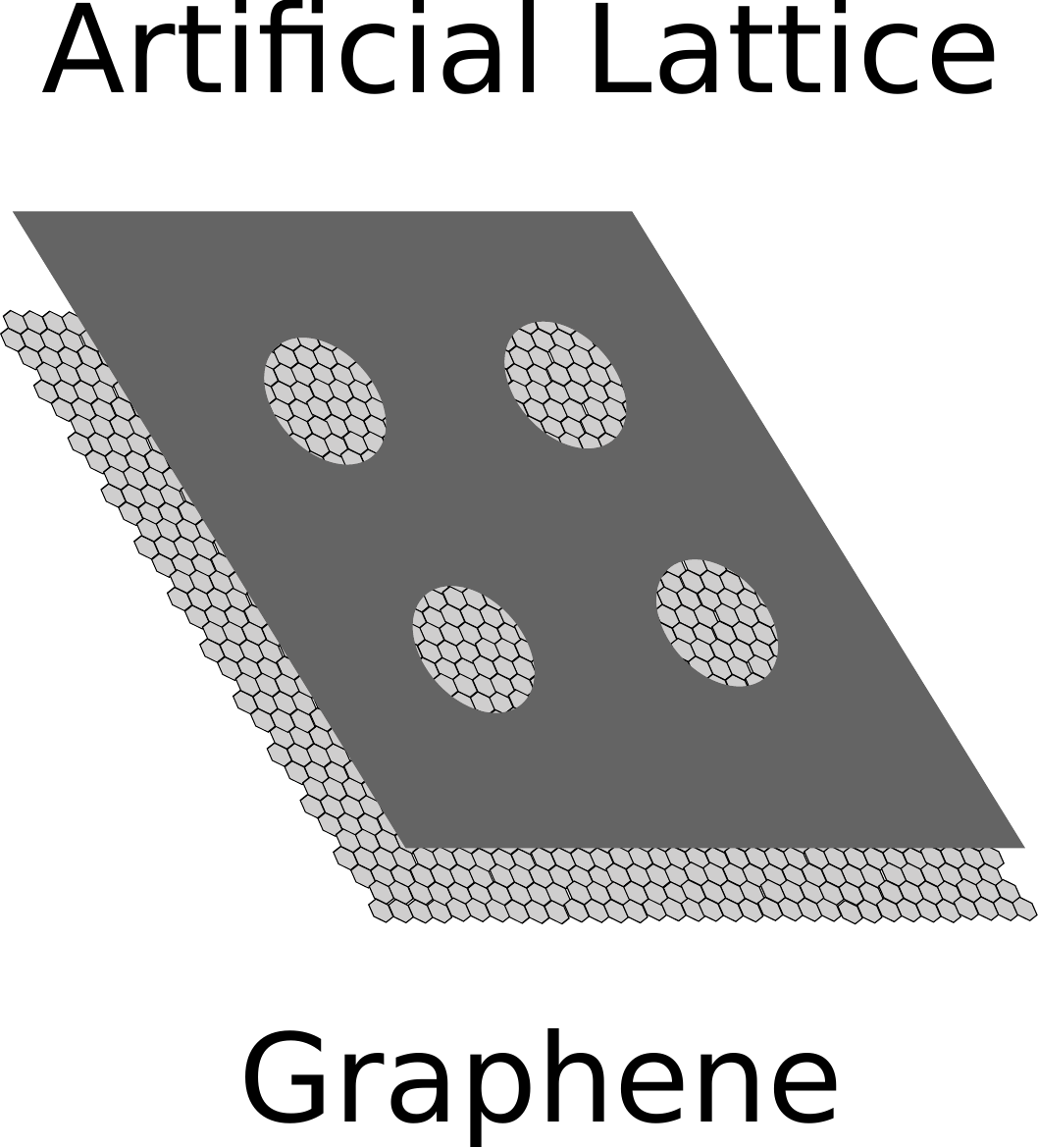} \quad \quad 
	\includegraphics[width=.47\columnwidth]{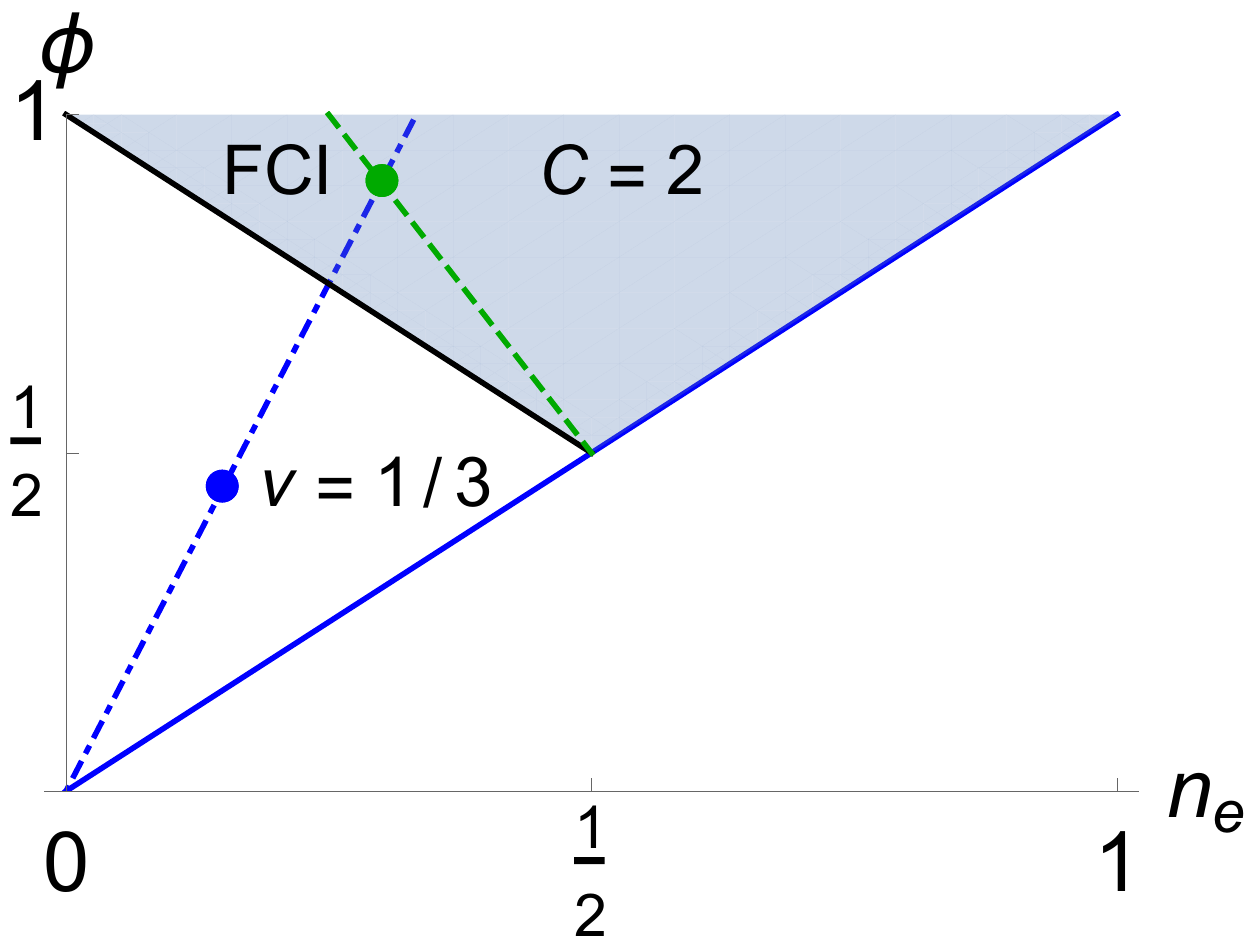}
	\caption{ {\it Left panel.}  The FCI can be engineered in graphene subject to an artificial lattice, {\it e.g.} by patterning holes in a neighboring dielectric or metal gate (see Appendix~\ref{app:experiments}). {\it Right panel.} Flux density versus electron density phase space.  The dot-dashed blue line corresponds to the FQH phase $\nu=1/3$.  The dashed green line corresponds to an FCI at quarter filling of a $C=2$, $S=-1$ band (shaded region).  The pair of points depict that for the different lattices shown in Fig.~\ref{fig:contacts}, the green and blue/yellow regions can be tuned to distinct phases for the same global backgate voltage and magnetic field.    }
	\label{fig:phase}
\end{center}
\end{figure}

%%%
{\it Lattice as a tuning parameter.}  We now focus on the particular realization of an FCI in graphene subject to an artificial lattice, depicted in the left panel of Fig.~\ref{fig:phase}.  Insulating phases correspond to lines in the flux density $\phi$ versus  electron density $n_e$ plane~\cite{Spanton17}.  The phase of the system can be tuned by: (1) applying a voltage to the sample to vary $n_e$, (2) applying a perpendicular magnetic field to vary $\phi$, and (3) changing the unit cell area of the lattice to change $(n_e,\phi)$ simultaneously.  
The third option provides a convenient way of realizing distinct phases within the same sample by defining the artificial lattice differently in separate spatial regions.  Here, we always consider edges defined by the artificial lattice, as the physical graphene edge is too dirty for translationally invariant physics to apply.

Consider the right panel of Fig.~\ref{fig:contacts}: the unit cell area in the green region is twice as large as the unit cell area in the blue/yellow regions.  Therefore, for the same magnetic field and backgate voltage, $2(n_e,\phi)_\text{blue/yellow}=(n_e,\phi)_\text{green}$.  When these points lie on lines characterizing distinct phases, the green region is in a different phase than the blue/yellow regions.  The right panel of Fig.~\ref{fig:phase} shows an example.  The dashed green line corresponds to an FCI at quarter filling of a $C=2,$ $S=-1$ band (shaded region).  A possible ground state of this phase is the Abelian $(331)$ state, which hosts genons at lattice dislocations.  The dot-dashed blue line corresponds to the FQH phase $\nu=1/3$.  
When the green region is tuned to the point $(3/10, 9/10)$, the blue/yellow regions are at $(3/20, 9/20)$.  Generally, for large $\phi$ FCI phases have larger energy gaps than competing FQH phases~\cite{Spanton17}, therefore for these parameter values we would expect the bulk and lattice contacts to be in an FCI and FQH phase, respectively.

%%%
{\it Layer-resolved lattice contacts}.
We now propose the layer-resolved lattice contacts shown in Fig.~\ref{fig:contacts}.  We assume the bulk region (green) is in the $(331)$ state, with parameter values given by the green dot in Fig.~\ref{fig:phase}.  We further assume the ground state is plaquette-centered $C_4$-symmetric, where the layer basis corresponds to the blue and yellow sublattices (see insets).  The two contacts (blue and yellow) are in the $\nu=1/3$ state corresponding to the blue dot in Fig.~\ref{fig:phase}, and are held at the same chemical potential. 
The FCI-contact interface is assumed to be long enough that translation symmetry is preserved, and located in the middle of the edge so that corner physics may be neglected. 

The white and green lines indicate the edge electrons $\psi_{e,1/2}$ associated with the FCI layer index.  These edge electrons are eigenstates of the translation operators $T_{(1,\pm 1)}$, and thus have well-defined momenta.  Due to the $C_4$ symmetry, the edge momentum of $\psi_{e,1/2}$ along the yellow contact interface is equal to the edge momentum of $\psi_{e,2/1}$ along the blue contact interface.  If the energy gaps of the $(331)$ and $\nu=1/3$ phases are compatible such that for an appropriate value of the chemical potential the lattice contact's edge electron has the same momentum as either $\psi_{e,1/2}$, then $\psi_{e,1}$ and $\psi_{e,2}$ can be gapped out along opposite contacts.  By tuning a global backgate, the electrochemical potential can be adjusted so that the contact's edge electron has the necessary momentum, which can be checked by the tuning procedure described in Appendix~\ref{app:experiments}.  Electron tunneling gaps out $\psi_{e,1/2}$ along the $\tau_{(1,\pm1)}=\pm\sigma_z$ invariant edges.
We do not show the edge electron associated with the filled $C=-1$ band (solid black line in Fig.~\ref{fig:phase}); generically this edge electron's momentum will be different than that of the $\psi_{e,1/2}$ and does not change under $C_4$ rotation, thus it can be safely ignored.
 Effectively, gapping out an FCI's edge electron along the contact's interface reroutes that edge electron along the exterior of the contact, spatially separating the FCI's two edge electrons.  A current measurement or voltage applied to the outer edge of the lattice contact will only affect one of the FCI's edge electrons, which is why we call these {\it layer-resolved lattice contacts}.   

\begin{figure}[t]
\begin{center}
	\includegraphics[width=0.7\columnwidth]{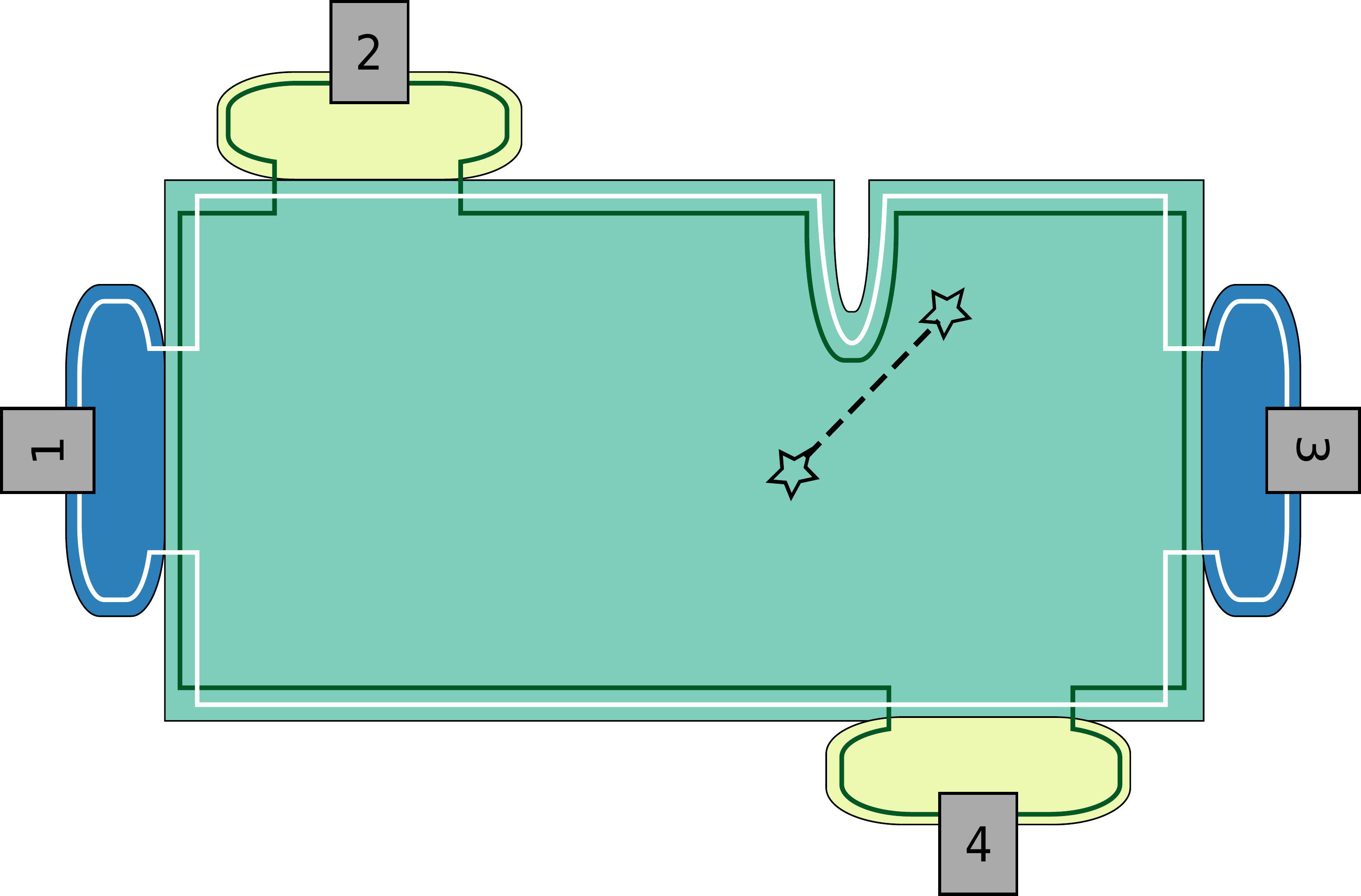}
	\caption{Detecting genons using FCI edges.  The two edge electrons (white and green lines) are interchanged at a genon (star), resulting in a signature in the differential conductance $dI_r/dV_r$~\cite{Barkeshli14}.  When all contacts are held to the same chemical potential, electrodes 1 and 3 selectively couple to one of the FCI's edge electrons, while 2 and 4 couple to the other.  By measuring the voltage drop between 1 and 3, as well as 2 and 4, we can determine the relative current $I_r$.  The differential conductance $dI_r/dV_r$ can then be determined by varying the voltage applied to any of the four electrodes.
}
	\label{fig:STM}
\end{center}
\end{figure}

Given the ability to separately contact the two edge electrons of an FCI, we can use their relative current to detect genons localized at lattice dislocations in the bulk.  Figure~\ref{fig:STM} generalizes one of the experimental proposals in Ref.~\onlinecite{Barkeshli14} for a conventional bilayer QH system with layer exchange defects.   Let $I_{1/2}$ denote the current associated with $\psi_{e,1/2}$.  The relative current $I_r=I_1-I_2$ is inverted across a genon.
The layer-resolved lattice contacts allow separate control of the voltage and measurement of the current for the two edge electrons, thereby allowing readout of their relative conductance, $dI_r/dV_r$.  The relative conductance peaks for small edge-genon separation; therefore by comparing multiple samples that vary this separation distance, we can obtain spatial resolution of the relative conductance and detect the genon.  
 The quantum point contact interferometer of Ref.~\onlinecite{Barkeshli14} can be similarly generalized to the FCI context.  

There are many other choices for the FCI and lattice contact phases; the two phases can be realized simultaneously for constant magnetic field and backgate voltage provided the line connecting ${(n_e,\phi)_\text{green}}$ and ${(n_e,\phi)_\text{blue/yellow}}$ intersects the origin.
For an FCI-contact interface along the $(1,0)$ direction with internal translation ${\tau_x=\sigma_x}$, the FCI edge electron operators $\psi_{e,1}\pm \psi_{e,2}$ are translation eigenstates.  By tuning a lattice contact to gap out the odd combination, the associated current $I_1-I_2$ can be measured on that contact, allowing for an alternate realization of the experiment depicted in Fig.~\ref{fig:STM}, see Appendix~\ref{app:experiments}.  Finally, while we focused here on a $C_4$-symmetric FCI, the proposal could be generalized to other lattices.

%%%
{\it Summary and Outlook}. In this work, we proposed layer-resolved lattice contacts for FCI edges.  The lattice contacts utilize the interplay of translation symmetry with internal component labels of the FCI state to selectively couple to one of the FCI's edge electrons.  Lattice contacts facilitate genon detection in the bulk by measuring the differential conductance associated with the relative current between the edge electrons, which in our proposal becomes a standard four terminal conductance measurement.  The experimental proposal in this paper could be realized using graphene subject to an artificial lattice.

For the example considered in this paper, the genons are MZMs; more exotic topological defects are possible for $(mml)$ phases with ${|m-l|>2}$~\cite{Barkeshli12}.  Important open questions include determining the energy gaps, ground states, and symmetries, of different fractionally filled Chern bands.   We have assumed there exist compatible phases for the FCI bulk and lattice contacts such that by tuning the electrochemical potential, the lattice contact's edge electron has the same momentum as one of the FCI's edge electrons; more detailed numerics are necessary to identify which candidate phases satisfy this propertry.  Additionally, the role of disorder, and whether it causes FCI edge modes to equilibrate, could affect the experiment proposed in Fig.~\ref{fig:STM}.  
More broadly, FCIs realized with an artificial lattice provide a playground for studying interfaces of different topological phases, including the transfer and sharing of information across the interface.

%%%
\section*{Acknowledgments}

C.K. acknowledges support from the NSF GRFP under Grant No. DGE $114085$.  E.M.S. and A.F.Y. were supported by the National Science Foundation under EAGER grant No. DMR-1836776.  E.M.S. acknowledges the support of the Elings Fellowship from the California NanoSystems Institute at the University of California, Santa Barbara. A.F.Y. acknowledges the support of the David and Lucile Packard Foundation.

\appendix

%%%
\section{Chern band as $|C|$-layer QH system} \label{app:mapping}

In this Appendix, we review how the single-particle orbitals of a Chern band characterized by $C$ and $S$ at rational flux density $\phi=p/q$ can be identified with Landau-gauge orbitals of a $|C|$-layer QH system at flux density $\bar{\phi}=\phi+S/C$~\cite{Qi11, Wu12, Barkeshli13,Harper14,Ippoliti18}.

Magnetic translations by lattice vectors do not commute due to the non-integer flux density:
\begin{equation}\label{eq:magnetic-algebra}
T_x T_y =e^{i2\pi \phi}T_y T_x,
\end{equation}
where $T_{x/y}=e^{i\left(k_{x/y}-e A_{x/y}/c\right)a}.$  
It is therefore useful to consider a magnetic unit cell (MUC) containing $p$ flux quanta, such that translations along the MUC, {\it e.g.}, $T_x, T_y^q$, do commute.  This choice of MUC corresponds to the gauge choice $\mathbf{A}=(0,Bx,0)$, for which translations act on the single-particle momentum states as 
\begin{align}
T_x \ket{k_x,k_y} &= e^{i k_x a} \ket{k_x,k_y}
\\ T_y \ket{k_x,k_y} &= e^{ik_y a}\ket{k_x +\phi G_x,k_y}.
\end{align}
In the above,  ${k_x \in [0, G_x)}$, ${k_y \in [0,G_y)}$, and ${G_x = 2\pi/a,}$ ${G_y=2\pi/qa}$.  

A partial Fourier transform on the momentum eigenstates results in Wannier orbitals
\begin{equation}\label{eq:Wannier}
\ket{k_x,b} = \int_{-G_y/2}^{G_y/2} \frac{d k_y}{\sqrt{2\pi}} e^{i k_y b q a} e^{i \varphi(k_x,k_y)}\ket{k_x,k_y},
\end{equation}
where $b$ denotes the $y$ coordinate.  Wannier orbitals satisfy twisted boundary conditions ${\ket{k_x+G_x,b}=\ket{k_x,b+C}}$.  There are several choices for the phase $\varphi(k_x,k_y)$~\cite{Wu12}, which will not be important for the present discussion.

A convenient basis change on Eq.~\eqref{eq:Wannier} allows us to map the single-particle orbitals to Landau-gauge orbitals of a $|C|$-layer QH system at effective flux density $\bar{\phi}$ and effective magnetic length $\bar{\ell}_B = a/\sqrt{2\pi\bar{\phi}^{} }$.  We denote the new basis by $\ket{\tilde{k}_x,\beta}$, where 
\begin{align}\label{eq:basis}
\tilde{k}_x &= k_x + G_x \frac{b}{C} \in \mathbb{R},
& \beta &= b- C \lfloor \frac{b}{C}\rfloor \in \mathbb{Z}_C.
\end{align}
Translations act on single particle states in this new basis as 
\begin{align}
T_x \ket{\tilde{k}_x,\beta} &= e^{i \tilde{k}_x a} e^{i 2\pi \beta/C} \ket{\tilde{k}_x,\beta}
\\ T_y \ket{\tilde{k}_x,\beta} &= \ket{\tilde{k}_x +\bar{\phi}G_x,\beta+S}.
\end{align}
Translations factor into a continuum and internal part, ${T_j=\tilde{T}_j\otimes \tau_j}$ for $j=x,y$.  We see that $\tilde{T}_j$ tranforms $\tilde{k}_x$ as in a continuum Landau level at flux density $\bar{\phi}$, while the layer index $\beta$ is acted on by 
\begin{align}\label{eq:tau-1}
\tau_x &= e^{-2\pi i \hat{\beta}/C}, & \tau_y &=\ket{\beta+S}\bra{\beta}.
\end{align} 

The square lattice additionally has a plaquette-centered ($p$) and site-centered ($s$) $C_4$ symmetry, related by $C_{4,s}=T_x C_{4,p}$.  Plaquette-centered $C_4$ symmetry satisfies $C_{4,p}^{-1}T_x C_{4,p}=T_y$ and $C_{4,p}^{-1}T_y C_{4,p}=T_x^{-1}$.    These rotations factor into continuum and internal parts, $\tilde{C}_{4,s/p} \otimes \gamma_{4,s/p}$, where the continuum part $\tilde{C}_{4,p}$ has the same action as $C_4$ acting on Landau-gauge orbitals in a continuum Landau level, while the internal part $\gamma_{4,p}$ acts on the layer indices as
\begin{align}
\gamma_{4,p} &= \sum_{\beta,\beta'} \frac{1}{\sqrt{C}} \ket{\beta}e^{2\pi i \beta \beta' /C}\bra{\beta'}.
\end{align}

%%% 
In the limit that ${a/\bar{\ell}_B \to 0}$, equivalently when $\phi\to -\frac{S}{C}$, the bands become flat and the system has a Landau level-like continuum limit.  Given any $|C|$-layer FQH state in this limit, an analogous FCI state is given by replacing the Landau-gauge orbitals with the single-particle basis states $\ket{\tilde{k}_x,\beta}$.  This construction is particularly simple on the cylinder, for which the $n_i$th Landau-gauge orbital sits at the $n_i$th lattice site in the compact direction. 
For example, on a cylinder compact in the $x$-direction with circumference $L_x$, the $(lmn)$ wavefunction is 
\begin{widetext}
\begin{align}
\Psi_{lmn} \{ z_i,w_j\} &= \Omega_z \Omega_w \prod_{i<j} \left( e^{2\pi z_i/L_x}-e^{2\pi z_j/L_x}\right)^l \left( e^{2\pi w_i/L_x}-e^{2\pi w_j/L_x}\right)^m \prod_{i,j} \left(e^{2\pi z_i/L_x}-e^{2\pi w_j/L_x}\right)^n e^{-\sum_i (y_i^2+v_i^2)/2\ell_B^2},
\end{align}
where $z_i=x_i+iy_i$ is the complex coordinate of the $i$th electron in the top layer, $w_j=u_j+iv_j$ is the complex coordinate of the $j$th electron in the bottom layer, and $\Omega_{z/w}$ are normalization factors.  We can rewrite the wavefunction in the occupation basis as 
\begin{align}
\Phi_{lmn}\{l_i,m_j\} &= \Omega_{l,m} \int \prod'_i \prod'_j dx_i dy_i du_j dv_j \psi^*_{\tilde{k}_{x,l_i}}(x_i,y_i) \psi^*_{\tilde{k}_{x,m_j}}(u_j, v_j) \Psi_{lmn}\{z_i,w_j\},
\end{align}
\end{widetext}
where $\Omega_{l,m}$ is a normalization factor, the primes on the products denote they are restricted to $\{i|l_i\in\{l_i\}\}$ and ${\{j|m_j\in \{m_j\}\}}$.  We have written the Landau-gauge wavefunction as $\psi^*_{\tilde{k}_x}$, and defined $\tilde{k}_{x,l_i}=G_x l_i/L_x$.  

The many-body state is then
\begin{align}\label{eq:mnl}
\ket{lmn}&= \sum_{\{l_i,m_j\}} \Phi_{lmn}\left(\{l_i,m_j\}\right)\prod'_i \prod'_j \ket{\tilde{k}_{x,l_i},0}\ket{\tilde{k}_{x,m_j},1}.
\end{align}
In writing Eq.~\eqref{eq:mnl}, we only relied on the fact that the continuum variable $\tilde{k}_x$ is analogous to momentum in a Landau level.  This fixes the definition of $\tilde{k}_x$, but we are free to rotate the basis of the layer index (the basis of the $\tau_j$) provided that $\tilde{k}_x$ is unaffected.  We refer to the {\it layer basis} as the single-particle basis choice in Eq.~\eqref{eq:mnl}, for which $\beta=0,1$ corresponds to the layer indices $l,m$ of the many-body state.  When the $\tau_j$ in the layer basis are given by Eq.~\eqref{eq:tau-1}, translations along $y$ permute the layers while translations along $x$ do not; this corresponds to the topological nematic states proposed in Ref.~\onlinecite{Barkeshli13}.   
We show below that if in the layer basis $\tau_x=\sigma_x$ and $\tau_y=\sigma_y$, then the $(mml)$ many-body ground state preserves $C_4$ symmetry.  Whereas for a QH system the different components correspond to different layers or spin species, for an FCI the components are associated with different lattice sites ({\it e.g.}, a checkerboard arrangement for the $C_4$-symmetric layer basis and a stripe arrangement for the topological nematic states).  

As the continuum translation operators $\tilde{T}_j$ act on the single particle basis in the same way as magnetic translations on Landau-gauge orbitals, we know that translations acting on the many-body state $\ket{lmn}$ can only differ in the actions of the internal part of translation $\tau_j$.  When $\tau_x=\sigma_z$ and $\tau_y=\sigma_x$ ({\it i.e.}, for the basis given in Eq.~\eqref{eq:tau-1} with $C=2$ and $S$ odd),
\begin{align}
&\tau_x \ket{lmn} = 
  (-1)^{|\{m_j\}|} \ket{lmn},
\\ &\tau_y \ket{lmn} =  
\ket{mln}.
\end{align}
In this basis, $\tau_y$ exchanges the layers, but $\tau_x$ does not: this many-body state does not preserve the microscopic $C_4$ symmetry of the lattice.

Conversely, if we rotate the component basis by the unitary $U = (1-i\sigma_y)(1-i\sigma_z)/2$, then $\tau_x=\sigma_x$ and $\tau_y=\sigma_y$.  When this corresponds to the layer basis, we have 
\begin{align}
\tau_x \ket{lmn} &= \ket{mln},
\\ \tau_y \ket{lmn} &=(-i)^{|\{l_i\}|} i^{|\{m_j\}|} \ket{mln}.
\end{align}
Assuming both layers have equal occupation, {\it i.e.} ${|\{l_i\}|=|\{m_j\}|}$, then both $\tau_x$ and $\tau_y$ exchange the layers.  When $l=m$, this many-body state is a $C_4$ eigenstate.  

%%%
\section{Translation symmetry for FCI edge states}\label{app:edge-details}

\begin{figure}[t]
\begin{center}
	\includegraphics[width=.9\columnwidth]{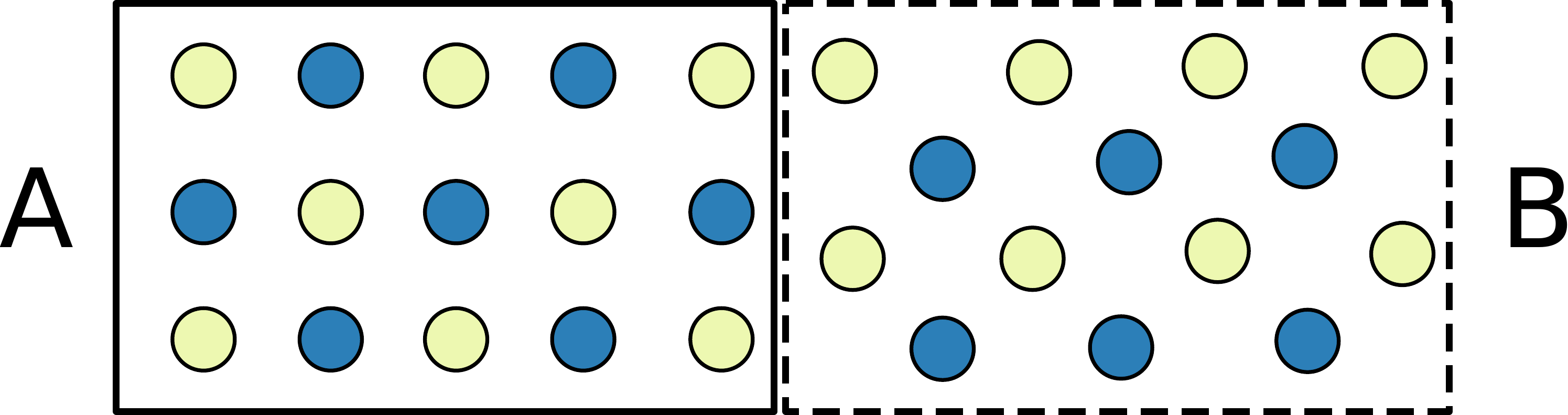}
	\caption{The left half of the sample has `A-type' edges, oriented along the $(1,0)$ and $(0,1)$ directions, with corresponding internal translations $\tau_x=\sigma_x$ and $\tau_y=\sigma_y$.  The right half has `B-type' edges, oriented along the $(1,1)$ and $(1,-1)$ directions, with internal translations $\tau_{x/y}=\pm \sigma_z$.  When translation symmetry is preserved along the edges, the interface has a gapless edge mode due to the lattice mismatch.  }
	\label{fig:interface}
\end{center}
\end{figure}

In this Appendix, we investigate the action of translation symmetry on edge states of an FCI in more detail.  For concreteness, we consider the $(mml)$ phase with the internal part of translation in the layer basis along the $(1,0)$ and $(0,1)$ directions given by $\sigma_x$ and $\sigma_y$, respectively.  We consider two types of edges, as shown in Fig.~\ref{fig:interface}: `A-type' edges run along the $(1,0)$ and $(0,1)$ directions, while `B-type' edges run along the $(1,1)$ and $(1,-1)$ directions.

More explicitly, the internal part of translation acts on the electron operators for A-type edges as 
\begin{align}
\sigma_x \left(\begin{array}{c}\psi_{e,1} \\ \psi_{e,2} \end{array}\right)&=  \left(\begin{array}{c}\psi_{e,2} \\ \psi_{e,1} \end{array}\right)
\\\sigma_y \left(\begin{array}{c}\psi_{e,1} \\ \psi_{e,2} \end{array}\right)&=  \left(\begin{array}{c}-i \psi_{e,2} \\ i \psi_{e,1} \end{array}\right),
\end{align}
and on the electron operators for B-type edges as 
\begin{align}
\pm \sigma_z \left(\begin{array}{c}\psi_{e,1} \\ \psi_{e,2}\end{array}\right)&=  \pm \left(\begin{array}{c}\psi_{e,1}\\ -\psi_{e,2} \end{array}\right).
\end{align}

In the main text, we argued that for symmetric edge potential $V_{IJ}=V_{JI}$ the edge electrons should have momentum difference $\pi/(\sqrt{2}a)$ along a B-type edge. 
Along an A-type edge with $\tau_z=\sigma_x$, the component translation eigenmodes are the symmetric and antisymmetric combinations of the electron operators, 
\begin{align}
T_{x/y} \left( \begin{array}{c} \psi_{e,1}+\psi_{e,2} \\ \psi_{e,1}-\psi_{e,2} \end{array}\right) &= e^{i \tilde{k}_{x/y}a} \left( \begin{array}{c} \psi_{e,1}+\psi_{e,2} \\ -\psi_{e,1}+\psi_{e,2} \end{array}\right),
\end{align}
indicating that the momentum separation of the edge electrons is $\pi/a$.  Therefore, when translation symmetry is preserved, an interface between an A-type and B-type edge of the same bulk phase, as depicted in Fig.~\ref{fig:interface}, cannot simultaneously gap out both edge electrons. 

We can check the momentum difference of the edge electrons explicitly for the bilayer checkerboard tight binding model~\cite{Yang12}.  This model has two bands with Chern numbers $C=\pm 2$, where each site on a square lattice has both an $a$ and a $b$ orbital, 
\begin{align}\label{eq:TBH}
H=t_1\sum_{\langle i,j\rangle}\left( e^{i\phi_{ij}} a_i^\dagger b_j +h.c\right) + \sum_{\langle \langle i,j \rangle \rangle} t_{ij} \left(a_i^\dagger a_j - b_i^\dagger b_j \right).
\end{align}
Equation~\eqref{eq:TBH} describes a bilayer checkerboard lattice.
We work with the parameters $\phi_x=\phi_{-x}=\pi/4$, ${\phi_y=\phi_{-y}=-\pi/4}$, and second nearest neighbor hoppings $t_{++}=t_{--}=t_2$, and $t_{+-}=t_{-+}=-t_2$.  On the  infinite plane, the Hamiltonian can be written in momentum space as 
\begin{align}
H&= \sum_{\vec{k}} \left( a_{\vec{k}}^\dagger, b_{\vec{k}}^\dagger\right) H_{\vec{k}} \left( \begin{array}{c} a_{\vec{k}} \\ b_{\vec{k}}\end{array}\right),
\\ H_{\vec{k}} &= \sqrt{2} t_1 \left( \cos k_x+\cos k_y \right) \sigma_x \notag
\\ &\quad  - \sqrt{2}t_1 \left(\cos k_x-\cos k_y \right)\sigma_y - 4 t_2 \sin k_x \sin k_y \sigma_z.
\end{align}

In order to investigate the edge modes, we can put the model on a strip finite in $x$ and infinite in $y$.  The Hamiltonian then is given by
\begin{align}\label{eq:H-straight}
H = \sum_{x=1}^n \sum_{k_y} \Big\{ t_1 a_{x,k_y}^\dagger   &\Big( e^{i\pi/4}b_{x+1,k_y} + e^{i\pi/4}b_{x-1,k_y} \notag
\\ &+ e^{-i \pi/4}2\cos k_y b_{x,k_y} \Big) + h.c.  \notag
\\ + t_2  2i\sin k_y \Big[& a_{x,k_y}^\dagger \left( a_{x+1,k_y}- a_{x-1,k_y} \right) \notag
\\ &- b_{x,k_y}^\dagger \left( b_{x+1,k_y}- b_{x-1,k_y} \right)\Big]\Big\} .
\end{align}

\begin{figure}[t]	\label{fig:TB}
	\includegraphics[width=0.495\columnwidth]{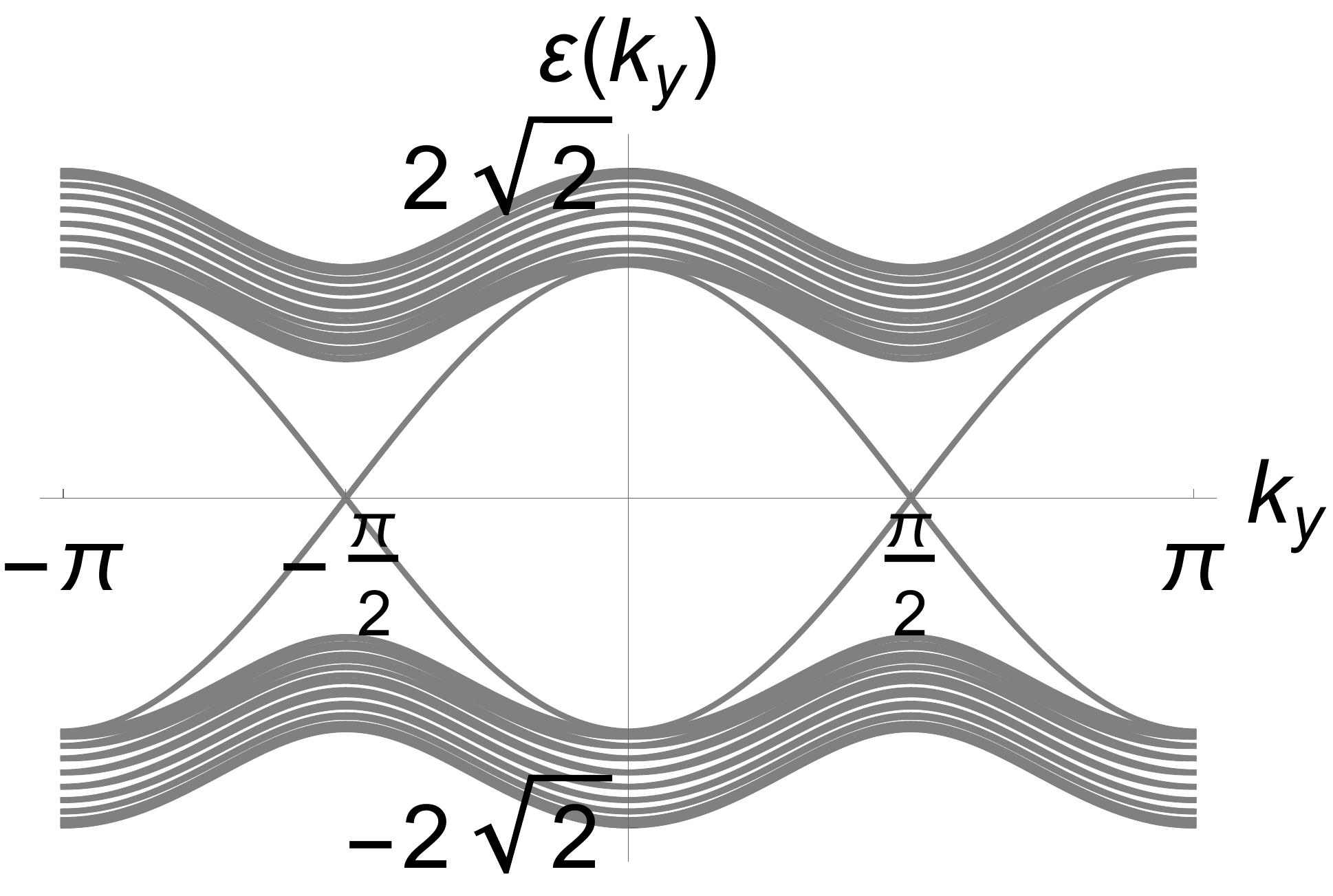}
	\includegraphics[width=0.495\columnwidth]{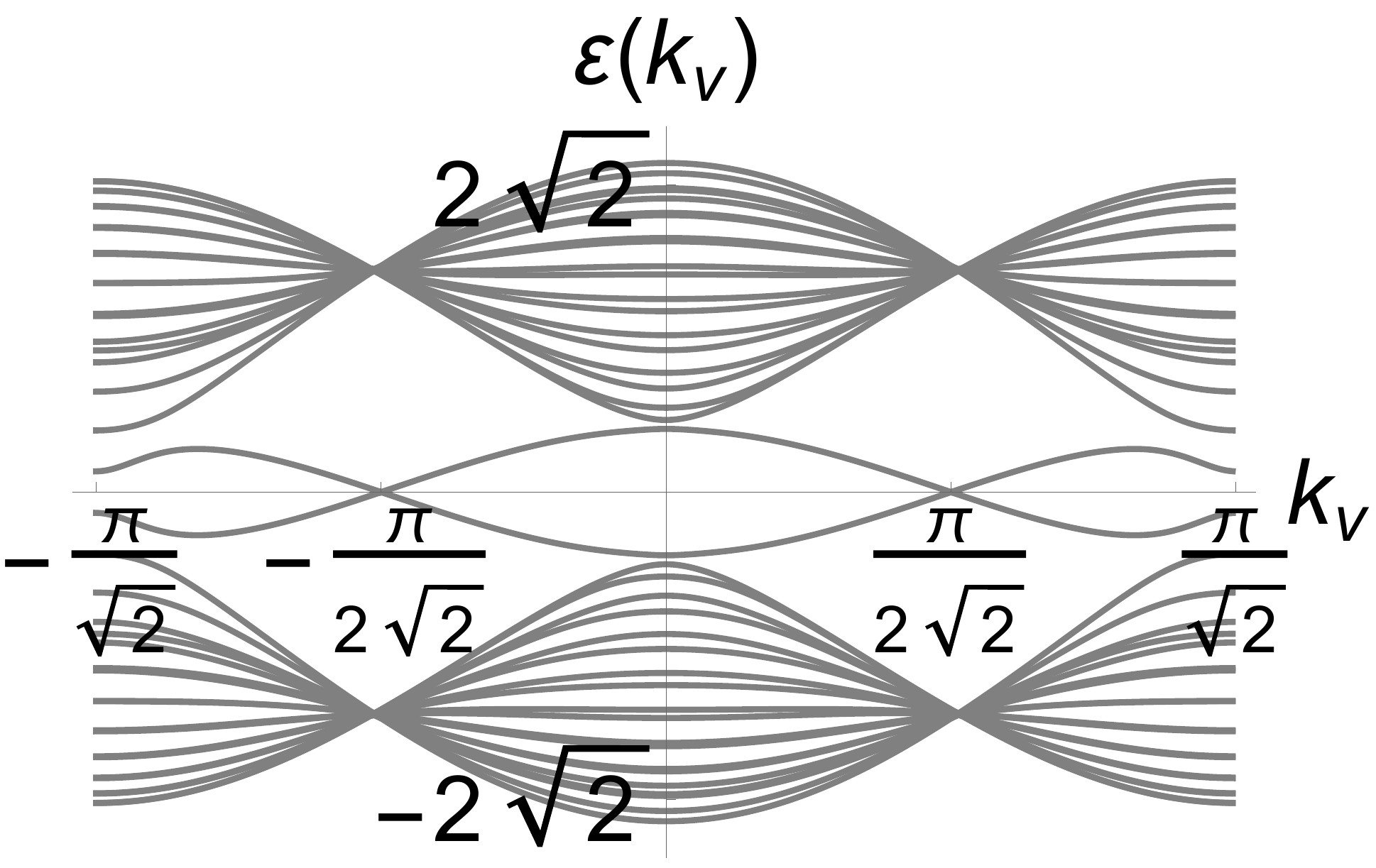}
	\caption{Energy versus momentum for the bilayer checkerboard lattice model on the infinite strip with edges parallel to the $y$-axis $(0,1)$ ({\it left panel}) and with edges parallel to the $v$-axis $(1,1)$ ({\it right panel}).  Right and left movers correspond to opposite edges.  The edge state momentum difference is quantized to $\pi/a$, where $a$ is the lattice spacing along the edge ({\it i.e.}, $a=1$ for the $y$ edge and $a=\sqrt{2}$ for the $v$ edge).  
 }
\end{figure}

We plot the energy spectrum of Eq.~\eqref{eq:H-straight} in the left panel of Fig.~\ref{fig:TB}.  As expected, the momentum difference of the edge modes is $\pi/a$.  Adding an orbital-dependent edge potential 
\begin{align}\label{eq:Ha}
H_{a}&=\sum_{x=1}^n \sum_{k_y}  \mu_a(x) a_{x,k_y}^\dagger a_{x,k_y},
\end{align}
affects both edge modes, therefore we conclude that the edge modes are mixtures of the two orbitals.  

Additionally, we can consider an infinite strip with B-type edges.  Denoting the $(1,1)$ coordinate by $v$ and the $(1,-1)$ coordinate by $u$, the Hamiltonian is given by 
\begin{align}\label{eq:H-diagonal}
&H= \sum_{u=1}^n \sum_{k_v} \notag
\\ & \Big\{  
 2 t_1\Big[  \cos \left( \frac{k_v}{\sqrt{2}}+ \frac{\pi}{4}\right)  \left( a_{u,k_v}^\dagger b_{u+\frac{1}{\sqrt{2}},k_v}+ b_{u,k_v}^\dagger a_{u-\frac{1}{\sqrt{2}},k_v} \right)  \notag
\\ &~~\quad + \cos\left( \frac{k_v}{\sqrt{2}}-\frac{\pi}{4}\right)  \left( a_{u,k_v}^\dagger b_{u-\frac{1}{\sqrt{2}},k_v} + b_{u,k_v}^\dagger a_{u+\frac{1}{\sqrt{2}},k_v}\right) \notag
\\ &+ t_2 \Big[a_{u,k_v}^\dagger \left( a_{u+\sqrt{2},k_v}+a_{u-\sqrt{2},k_v} - 2 \cos (\sqrt{2} k_v) a_{u,k_v} \right) \notag
\\ &~~\quad -b_{u,k_v}^\dagger \left( b_{u+\sqrt{2},k_v}+b_{u-\sqrt{2},k_v}  - 2 \cos (\sqrt{2} k_v) b_{u,k_v} \right) \Big]
\Big\}.
\end{align}

We plot the energy spectrum of Eq.~\eqref{eq:H-diagonal} in the right panel of Fig.~\ref{fig:TB}.  Now, the lattice spacing along the edge is $\sqrt{2}a$, and we see that the momentum difference of the edge modes is $\pi/\sqrt{2}a$.  Adding the orbital-dependent edge potential of Eq.~\eqref{eq:Ha} only affects one left-moving and one-right moving mode, thus we conclude that each edge mode can be identified with a single orbital.

For these parameter values, the flux is tuned exactly to ${\phi=1/2=-S/C}$, which is why the edge momentum difference is well quantized to $\pi/a$ ($\pi/\sqrt{2}a$) for A-type (B-type) edges.  We now further investigate the edge momentum difference as $\phi\to -S/C$ for the Hofstadter model~\cite{Hofstadter76,Harper55}. 

\begin{figure}[t]	
	\includegraphics[width=\columnwidth]{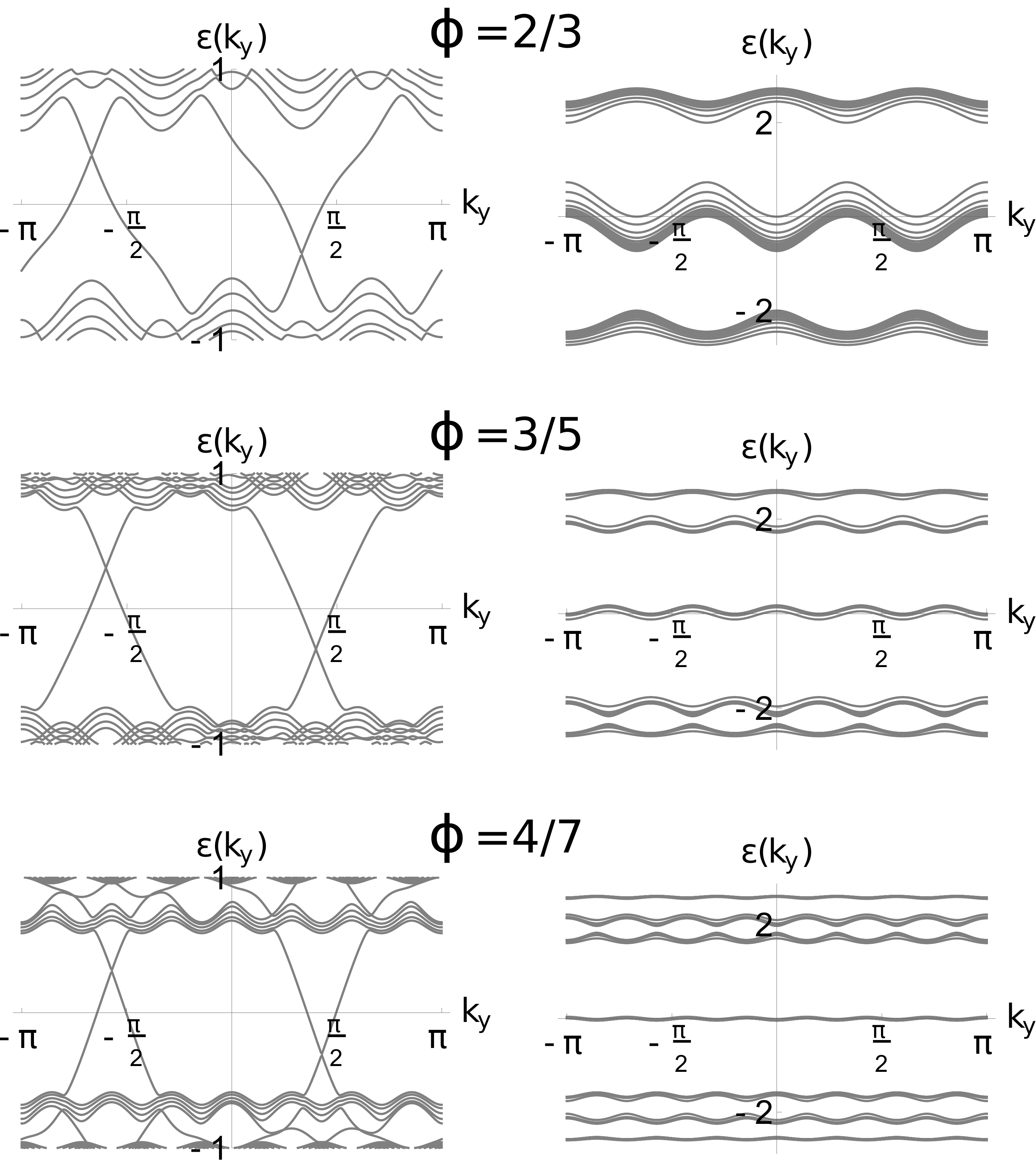}
	\caption{Energy spectrum for the Hofstadter model near $\phi=1/2$. {\it Left panel.} Edge state momentum difference for the same setup as described in Fig.~\ref{fig:Hof1} at $\phi=2/3$, $3/5,$ and $4/7$. {\it Right panel.}  Energy plotted against $k_y$ for different values of $k_x$ for the infinite plane. As $\phi\to 1/2$, the middle band becomes flatter. }\label{fig:HofModel}
\end{figure}

The Hofstadter model describes a perpendicular magnetic field applied to a square lattice (with lattice constant $a$):
\begin{align}
H= -t \sum_{\langle m,n\rangle} \left\{ c_n^\dagger c_m e^{i 2\pi \int_m^n \vec{A}\cdot d\vec{\ell}}+h.c.\right\}.
\end{align}
Working in Landau gauge $\vec{A}=(0,Bx,0)$ at flux density $Ba^2=\phi=p/q$, we can write the Hamiltonian in momentum space on the infinite plane as 
\begin{align}
H= \sum_{\vec{k}} \vec{c}_{\vec{k}}^\dagger \mathcal{H}_{\vec{k}} \vec{c}_{\vec{k}},
\end{align}
where $\vec{c}_{\vec{k}}=\left(c_{\vec{k},1},c_{\vec{k},2},\dots, c_{\vec{k},q}\right)^T$ and $\mathcal{H}(\vec{k})$ is a $q\times q$ matrix, defined by
\begin{align}
\mathcal{H}(\vec{k})_{m,n}= \cos\left(k_y a +2\pi n \phi\right)\delta_{m,n}+e^{i\frac{ k_x a}{q}} \delta_{m+1,n}+h.c.
\end{align}
The magnetic unit cell has $q$ sites in the $y$ direction, one site in the $x$ direction.  Near $\phi=1/2$, energy spectrum has three bands, with the middle band having $C=2,S=-1$.  By placing the model on a cylinder infinite in $x$, such that half of the $y$ sites sit at chemical potential $\mu_1$ between the bottom two bands and half of the $y$ sites sit at chemical potential $\mu_2$ between the top two bands, we can investigate the edge states of the Chern $2$ band.  We plot the energy spectrum of the infinite plane and for the infinite cylinder in Fig.~\ref{fig:HofModel}, at flux densities $\phi=p/q=1/2+m/n$ for $p/q=\{2/3,3/5,4/7\}$ and $m/n=\{1/6,1/10,1/14\}$.  We see that as $\phi\to 1/2$, the Chern-2 band becomes flatter, and the difference in edge state momenta becomes closer to $\pi/a$.  At zero energy, the differences in edge momenta are $\pi/a(1+1/3), \, \pi/a(1+1/5),$ and $\pi/a(1+1/7)$, respectively.  Therefore, we see as $\phi\to -S/C$, the edge state momentum difference approaches quantization at $\pi/a$.  This agrees with the perturbation theory discussion of the Hofstadter model in Ref.~\onlinecite{Harper14}.

%%%
\section{Corner MZMs}\label{app:MZMs}

In this Appendix, we discuss corner MZMs in a $C_4$-symmetric $(331)$ state.  A similar argument holds for $(mml)$ states with $|m-l|=2$.  The edge Lagrangian written in terms of the charged ($+$) and neutral ($-$) boundary modes  ${\psi_\pm\sim e^{i\left(\phi_1\pm \phi_2\right)}}$ is
\begin{equation}
L_0 = \int dx \left\{ v_- \psi_- \partial_x \psi_- + v_+ \psi_+ \partial_x \psi_+\right\}.
\end{equation}
Note that $\psi_-$ is a fermion (has scaling dimension $1/2$).
The component plaquette-centered component $C_4$ symmetry ${\gamma_{4,p}=\left(\sigma_x+\sigma_y\right)/\sqrt{2}}$ implies that the edge modes transform as  
\begin{align}\label{eq:gamma4p}
\gamma_{4,p} \left( \begin{array}{c} \phi_1 \\ \phi_2 \end{array}\right) &=  \left( \begin{array}{c} \phi_2 - \frac{\pi}{8} +2\pi b \\ \phi_1 + \frac{\pi}{8} + 2\pi c \end{array}\right).
\end{align}
The constants ${b= (3j-k)/8}$ and ${c=(3k-j)/8}$ for ${j,k\in \mathbb{Z}}$ come from the compactness of the fields ${\phi_{1/2}(x)+2\pi=\phi_{1/2}(x)}$.  Different values of $b$ and $c$ correspond to different symmetry fractionalization classes~\cite{Cheng16}.   

The boundary Lagrangian~\cite{Cardy04,Fendley09} that enforces the $C_4$ symmetry at a corner ({\it e.g.}, the Northwest corner at $z=0$) is 
\begin{align}\label{eq:boundary-lagrangian}
L_b =& -v_-\left\{  e^{-i\alpha_-} \psi_-^\text{W}(0)\psi_-^\text{N}(0) + e^{i \alpha_-} \bar{\psi}_-^\text{W}(0)\bar{\psi}_-^\text{N}(0)\right\} \notag
\\ &-v_+ \left\{ e^{-i\alpha_+} \bar{\psi}_+^\text{W}(0) \psi_+^\text{N}(0) +e^{i\alpha_+} \psi_+^\text{W}(0) \bar{\psi}_+^\text{N}(0)\right\}.
\end{align}
The phase factors $\alpha_\pm$ are defined by Eq.~\eqref{eq:gamma4p}, and are unimportant for the present discussion.  Equation~\eqref{eq:boundary-lagrangian} describes Andreev reflection of the neutral edge electron $\psi_-$, indicating Majorana zero modes (MZMs) at the corner of the system.
For ${b=c=0}$, $\psi_+$ is invariant under $\gamma_{4,p}$ (the charged mode is insensitive to the MZM), while $\psi_-$ transforms as ${\gamma_{4,p} \psi_- \to \bar{\psi}_- e^{-\frac{\pi}{4}}}$.  

Alternatively, when the system satisfies site-centered $C_4$ symmetry, then 
\begin{align}\label{eq:gamma4s}
\gamma_{4,s} \left( \begin{array}{c} \phi_1 \\ \phi_2 \end{array}\right) &=  \left( \begin{array}{c} \phi_1 + \frac{\pi}{8} +2\pi b \\ \phi_2 -\frac{\pi}{8} + 2\pi c \end{array}\right).
\end{align}
In particular, the rotation symmetry does not interchange the layers.  The boundary Lagrangian that enforces this symmetry is 
\begin{align}
L_b =& -v_-\left\{  e^{-i\alpha_-}\bar{\psi}_-^\text{W}(0)\psi_-^\text{N}(0) + e^{i \alpha_-} \psi_-^\text{W}(0)\bar{\psi}_-^\text{N}(0)\right\} \notag
\\ &-v_+ \left\{ e^{-i\alpha_+} \bar{\psi}_+^\text{W}(0) \psi_+^\text{N}(0) +e^{i\alpha_+} \psi_+^\text{W}(0) \bar{\psi}_+^\text{N}(0)\right\},
\end{align} 
where $\alpha_\pm$ are redefined compared to Eq.~\eqref{eq:boundary-lagrangian}.  In this case, both $\psi_\pm$ satisfy normal reflection, therefore there is no corner MZM.

Finally, we note that a system with plaquette-centered $C_4$ symmetry with A-type edges should not be interpreted as having corner MZMs, as all translation interchange the edge components and thus there is nothing special about the corner compared to the middle of the edge.  More explicitly, we could write a boundary Lagrangian to enforce the translation symmetry at the midpoint of a $(1,0)$ or $(0,1)$ edge.  Comparing
\begin{align} \label{eq:taux}
\sigma_x \left( \begin{array}{c} \phi_1 \\ \phi_2 \end{array}\right) &= \left(\begin{array}{c} \phi_2 +2\pi  b \\ \phi_1 +2\pi c  \end{array} \right)
\\ \sigma_y \left( \begin{array}{c} \phi_1 \\ \phi_2 \end{array}\right) &= \left(\begin{array}{c} \phi_2- \frac{\pi}{4}+ 2\pi b \\ \phi_1 + \frac{\pi}{4} +2\pi c  \end{array} \right) \label{eq:tauy},
\end{align}
to Eq~\eqref{eq:gamma4p}, we see this boundary Lagrangian has the same form as Eq.~\eqref{eq:boundary-lagrangian}, up to a redefiniton of the phases $\alpha_\pm$.  An equivalent statement is that in the case of B-type edges with plaquette-centered $C_4$ symmetry, there is a boundary entropy of $\ln\sqrt{2}$ associated with the corners that is not present mid-edge.  For A-type edges, the corners and mid-edges have the same boundary entropy.  We only identify a corner MZM when there is a $\ln\sqrt{2}$ boundary entropy difference between a mid-edge point and the corner~\cite{Affleck91,Fendley06,Fendley09}.

%%%
\section{Experiment}\label{app:experiments}

In this Appendix, we provide more details of the experimental proposal and additional tests of the layer-resolved lattice contacts.  

\begin{figure}[t]
	\includegraphics[width=\columnwidth]{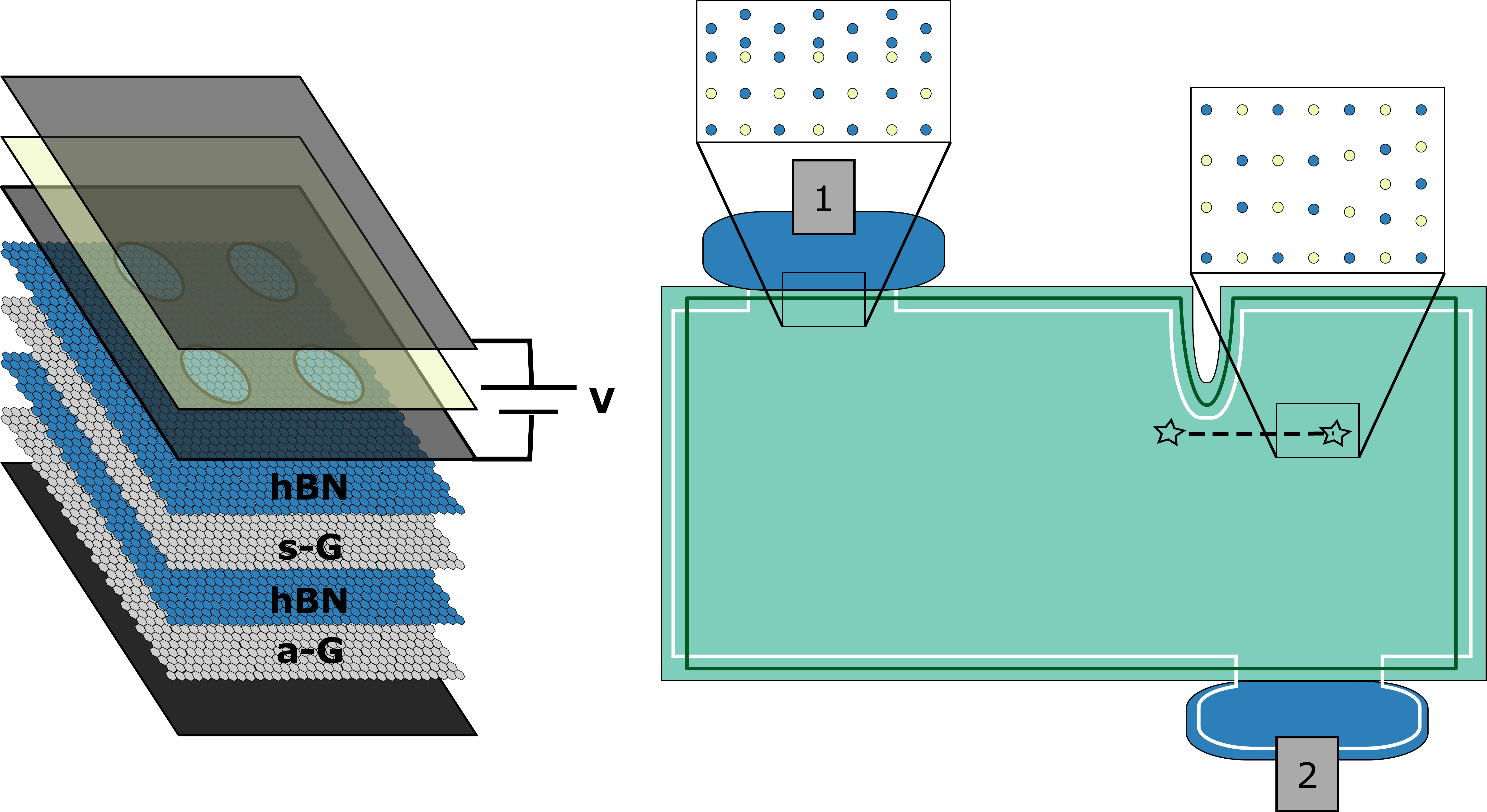}
	\caption{ {\it Left panel.} Artificial lattice proposal: a dielectric layer (yellow) separates two metal gates (gray) held at a potential difference $V$.  The bottom metal gate has a square lattice patterned into it, which applies a periodic potential to all lower layers.  A sacrificial graphene layer (s-G) is used to screen noise from the metal gates, so that only the periodic potential from the artificial lattice is applied to the active graphene layer (a-G) hosting the FCI.  The graphene layers are sandwiched between hexagonal Boron Nitride (h-BN), and a graphite bottom gate (black) varies the electron density of the sample. 
{\it Right panel.}  The same experiment as in Fig.~\ref{fig:STM}, using lattice contacts for A-type edges.  The white line now depicts the edge electron associated with $\psi_{e,1}-\psi_{e,2}$, which can be gapped out by appropriately tuning the chemical potential of the contacts.  The current associated with this edge electron is the relative current $I_r=I_1-I_2$.}
	\label{fig:artificial-lattice}
\end{figure}

The left panel of Fig.~\ref{fig:artificial-lattice} illustrates one approach to engineering an artificial lattice applied to graphene.  The lattice is patterned into a metal gate (third layer).  The potential from this lattice can be made larger by increasing the voltage difference between two metal gates, separated by a dielectric (yellow).   A sacrificial graphene layer (s-G) screens the metal gates to avoid introducing additional noise.   For large enough potential difference $V$, only the periodic potential from the artificial lattice passes through the screening layer to affect the active graphene layer (a-G).  The electron density of the system can be tuned using a graphite bottom gate (black).  

In order to ascertain that the layer-resolved lattice contacts are working as intended, we propose the following calibration experiments.  Each experiment could be first done with the bulk and contacts in filled $C=2$ and $C=1$ bands, respectively, then applied to the fractional case discussed in the main text.  
Here, we assume that in both cases the energy gaps of the bulk and contacts are compatible so that the relative chemical potential can be used to tune the lattice contact's edge momentum equal to the momentum of either of the bulk's edge electrons.
\\

\noindent{\it Testing interface length}. The layer-resolved lattice contacts rely on the contact-bulk interface being long enough so that translation symmetry is preserved.  To test this lengthscale, two bulk phases can be engineered such that their lattices are rotated by $\pi/4$ from each other, corresponding to an interface between an A-type and a B-type edge, as depicted in Fig.~\ref{fig:interface}.  When translation symmetry is preserved, there is a gapless edge mode running along the interface, since the two phases have different momentum separation.  When the interface is too short to preserve translation symmetry, the interface should be gapped out since the lattice orientation is unimportant for the physics.   
\\ 

\noindent{\it Testing separation from corners}.  The next test is for the setup shown in Fig.~\ref{fig:contacts}: all edges are B-type, with the West and South edges corresponding to blue and yellow sublattices, respectively.  By changing the displacement of the lattice contacts from the Southwest corner (by using different samples), we can check that the corner physics is not affecting the lattice contacts. When the lattice contact is far away from the corners, the momentum difference between the two edge states should be $\pi/(\sqrt{2}a)$.  When the contact is at the corner, there is no momentum difference between the two edge states.  By varying the lattice contact displacement from the edge and measuring the amount by which we need to tune the lattice contact's chemical potential to change from gapping out one edge mode to gapping out the other, we can measure the length scale over which the corner physics is important.  Once the lattice contacts are far-enough separated from the corner, the momentum difference between edge states will be fixed. A potential difficulty of this experiment is that we need to compare different samples, thus this calibration will only work if the edge potential does not vary strongly between samples.  Up to four displacements can be tested on the same sample by placing lattice contacts on different edges.  
\\

\noindent {\it Testing layer-selectivity}.  The final calibration test is again for the setup in Fig.~\ref{fig:contacts}, where one lattice contact controls voltage and the other measures current.  When the contacts are equilibrated with the FCI edge, the current injected at the contact should be equal to the voltage measured at that edge multiplied by the expected Hall conductance.  For a long contact, this should only happen when the contact's and FCI's edge electrons have the same momentum.
When the system is tuned so that the contacts gap out opposite edge electrons, varying $V$ on the blue contact should not affect the $I$ measured on the yellow contact.  When the contacts are tuned such that they should gap out the same edge electron, varying $V$ on the blue contact should directly affect the $I$ measured on the yellow contact.  The first case corresponds to the lattice contacts held at the same electrochemical potential  for the geometry shown (South and West edges corresponding to opposite sublattices).  The electrochemical potential of one contact can then be tuned (using an additional gate) to achieve the second case.
\\

\noindent{ \it Lattice contacts for A-type edges.}
The layer-resolved lattice contacts proposed in the main text use B-type edges for the FCI.  Alternatively, we can design lattice contacts for A-type edges for the FCI, as shown in the right panel of Fig.~\ref{fig:artificial-lattice}.  From Appendix~\ref{app:edge-details}, we know that the edge translation eigenstates are $\psi_{e,\pm}=\psi_{e,1}\pm \psi_{e,2}$ for $\tau_x=\sigma_x$.  When the contact is tuned to gap out $\psi_{e,-}$, a current measurement gives the exciton current $I_r$, which should carry the signature of the genon.  It is still important in this case that the FCI-contact interface is long enough that translation symmetry is preserved, and that the contact is unaffected by corner physics. 
\\

\noindent {\it Short contacts.}
When the FCI-contact interface is too short to preserve translation symmetry, we can still selectively couple to the different FCI layers of an $(lmn)$ phase with $l\neq m$ by using the lattice to tune the contacts into different phases.  In this case, the allowed perturbations at the interface are subject to the same conditions as for conventional QH systems. Short contacts can also be used when the ground state of the system spontaneously breaks the microscopic symmetry of the lattice.

%%%
%\nocite{apsrev41Control}
%\bibliographystyle{apsrev4-1}
%\bibliography{FCI-bib}


\end{document}